\documentclass[10pt,english]{article}
\usepackage[T1]{fontenc}
\usepackage{geometry}
\usepackage{babel}
\usepackage{verbatim}
\usepackage{mathrsfs}
\usepackage{amsmath}
\usepackage{amssymb}
\usepackage{graphicx}
\usepackage[unicode=true,
 bookmarks=true,bookmarksnumbered=true,bookmarksopen=true,bookmarksopenlevel=1,
 breaklinks=false,pdfborder={0 0 0},backref=false,colorlinks=false]
 {hyperref}
\hypersetup{pdftitle={Your Title},
 pdfauthor={Your Name},
 pdfpagelayout=OneColumn, pdfnewwindow=true, pdfstartview=XYZ, plainpages=false}

\makeatletter

\providecommand{\tabularnewline}{\\}

\newtheorem{assumption}{Assumption}
\newtheorem{problem}{Problem}
\newtheorem{theorem}{Theorem}
\newtheorem{definition}{Definition}

\date{}

\makeatother

\begin{document}

\title{Direct design of LPV feedback controllers:\\
technical details and numerical examples}

\author{Carlo Novara %
\thanks{Dipartimento di Automatica e Informatica, Politecnico di Torino, Italy.
E-mail: \protect\href{http://carlo.novara@polito.it}{carlo.novara@polito.it}.%
}}
\maketitle
\begin{abstract}
The paper contains technical details of recent results developed by
the author, regarding the design of LPV controllers directly from
experimental data. Two numerical examples are also presented, about
control of the Duffing oscillator and control of a two-degree-of-freedom
manipulator.

\end{abstract}

\section{Introduction}

Consider a discrete-time LPV (Linear Parameter-Varying) system $S$
in state-space form:
\begin{equation}
x_{t+1}=A\left(p_{t}\right)x_{t}+B\left(p_{t}\right)u_{t}+H\left(p_{t}\right)e_{t}\label{ss_sys}
\end{equation}
where $t\in\mathbb{Z}$ is the time, $x_{t}\in X\subseteq\mathbb{R}^{n_{x}}$
is the state, $p_{t}\in P\subset\mathbb{R}^{n_{p}}$ is the time-varying
parameter, $u_{t}\in U\subseteq\mathbb{R}$ is the input, and $e_{t}\in E\subseteq\mathbb{R}^{n_{e}}$
is a noise including both process and measurement disturbances (see
the Appendix). $P$ and $E$ are compact sets. $A$, $B$ and $H$
are matrices/vectors of suitable dimensions. $B$ and $H$ are Lipschitz
continuous functions of $p\in P$.

This manuscript contains technical details of recent results developed
by the author, regarding the direct design of LPV controllers for
systems of the form \eqref{ss_sys} from experimental data. Two numerical
examples are also presented: The first one regards control of the
Duffing oscillator, the second one control of a two-degree-of-freedom
robot manipulator.

\section{Problem formulation}

\label{prob_form}

Suppose that the matrices $A$, $B$ and $H$ in (\ref{ss_sys}) are
unknown, but a set of measurements is available:
\begin{equation}
\mathcal{D}\doteq\left\{ \widetilde{p}_{k},\widetilde{x}_{k},\widetilde{u}_{k}\right\} _{k=-L}^{-1}\label{D_inv}
\end{equation}
where $\widetilde{x}_{k}=x_{k}$ is the measured state (including
process and measurement noises), $\widetilde{u}_{k}$ is the noise-corrupted
measurement of $u_{k}$ and $\widetilde{p}_{k}=p_{k}$ is the measured
scheduling vector. For simplicity but without loss of generality,
$p_{k}$ is assumed to be noise-free (supposing a noise-corrupted
$p_{k}$ would lead to the same results, at the expense of an heavier
notation). $\widetilde{p}_{k}$, $\widetilde{x}_{k}$ and $\widetilde{u}_{k}$
are assumed bounded for all $k=-L,\ldots,-1$.

\begin{figure}[h]
\begin{centering}
\includegraphics[scale=0.5]{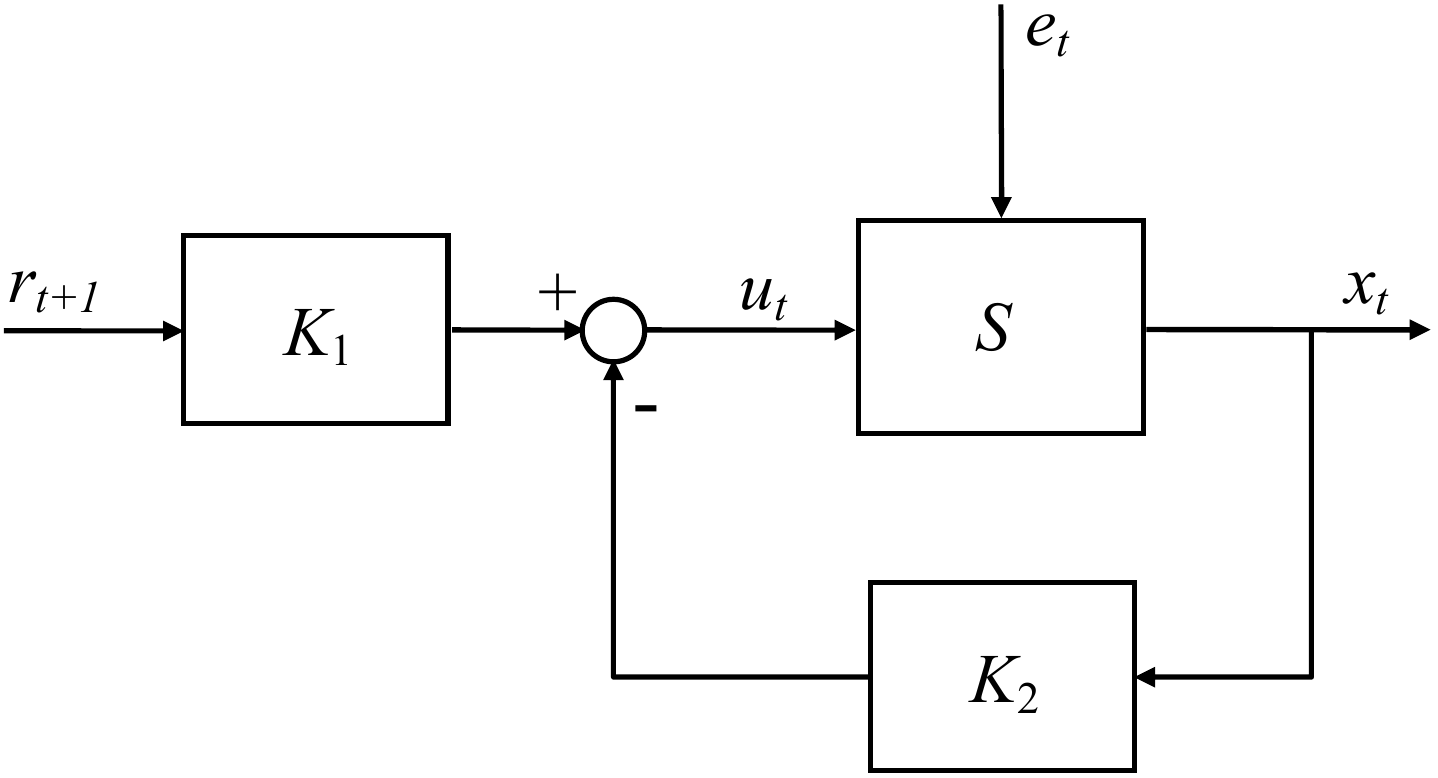} 
\par\end{centering}

\protect\caption{Feedback control system.}

\label{ctr_sys_1} 
\end{figure}

Consider the control system of Figure \ref{ctr_sys_1}, where $S$
is the LPV system (\ref{ss_sys}) and the controllers $K_{1}$ and
$K_{2}$ are vectors of the form 
\[
\begin{array}{c}
K_{1}\left(p_{t}\right)=(K_{11}\left(p_{t}\right),\ldots,K_{1n_{x}}\left(p_{t}\right))\\
K_{2}\left(p_{t}\right)=(K_{21}\left(p_{t}\right),\ldots,K_{2n_{x}}\left(p_{t}\right)).
\end{array}
\]
These vectors are functions of the scheduling parameter $p_{t}$ ;
the output of the block $K_{1}$ in Figure \ref{ctr_sys_1} is given
by $K_{1}\left(p_{t}\right)r_{t+1}$; the output of the block $K_{2}$
is given by $K_{2}\left(p_{t}\right)x_{t}$. The notation $Kx$ is
used to indicate the dot product between the vectors $K$ and $x$.

The problem is to design a controller $K\left(p_{t}\right)\doteq(K_{1}\left(p_{t}\right),K_{2}\left(p_{t}\right))$
ensuring a \textquotedbl{}small\textquotedbl{} tracking error $\left\Vert r_{t}-x_{t}\right\Vert _{\infty}$.
A \textquotedbl{}low\textquotedbl{} complexity controller is looked
for, allowing an efficient on-line implementation in real-world applications.

In order to formulate more precisely this problem, define the following
\emph{point-wise inversion error}:
\begin{equation}
\begin{array}{c}
IE\left(K,p,r,x,e\right)\doteq\left\Vert r-\left(A\left(p\right)-B\left(p\right)K_{2}\left(p\right)\right)x\right.\\
\left.-B\left(p\right)K_{1}\left(p\right)r-H\left(p\right)e\right\Vert _{\infty}
\end{array}\label{eq:ie_def}
\end{equation}
where $p\in P$, $(r,x)\in X^{2}$, $e\in E$, and $\left\Vert \cdot\right\Vert _{\infty}$
is the vector $\ell_{\infty}$ norm. Thus, define \emph{global inversion
error} as
\[
GIE\left(K\right)\doteq\left\Vert IE\left(K,\cdot,\cdot,\cdot,\cdot\right)\right\Vert _{L_{\infty}}
\]
where $\left\Vert \cdot\right\Vert _{L_{\infty}}$ is the functional
$L_{\infty}$ norm evaluated over the $IE$ domain $P\times X^{2}\times E$.

The following stability notion can also be introduced (for simplicity
but without loss of generality, zero initial conditions are assumed
in the remainder of the paper). \smallskip{}

\begin{definition}An LPV system with state $x_{t}$, input $u_{t}$,
parameter $p_{t}$, and noise $e_{t}$ is input-to-state $\ell_{\infty}$
stable if, some $\lambda_{u}<\infty$ and $\lambda_{e}<\infty$ exist
such that the state sequence $\mathbf{x}=(x_{1},x_{2},\ldots)$ is
bounded as 
\[
\left\Vert \mathbf{x}\right\Vert _{\infty}\leq\lambda_{u}\left\Vert \mathbf{u}\right\Vert _{\infty}+\lambda_{e}\left\Vert \mathbf{e}\right\Vert _{\infty}
\]
for any input sequence $\mathbf{u}=(u_{1},u_{2},\ldots)\in\ell_{\infty}$,
any noise sequence $\mathbf{e}=(e_{1},e_{2},\ldots)\in\ell_{\infty}$,
any parameter sequence $\mathbf{p}=(p_{1},p_{2},\ldots)\in\ell_{\infty}$
with $p_{t}\in P$, $\forall t$.$\qquad\blacksquare$\end{definition}

\smallskip{}

Consider now that the closed-loop system in Figure \ref{ctr_sys_1}
is described by the difference equation
\begin{equation}
\begin{array}{c}
x_{t+1}=\left(A\left(p_{t}\right)-B\left(p_{t}\right)K_{2}\left(p_{t}\right)\right)x_{t}\qquad\\
+B\left(p_{t}\right)K_{1}\left(p_{t}\right)r_{t+1}+H\left(p_{t}\right)e_{t}.
\end{array}\label{eq:cl_sys}
\end{equation}
Stabilizability is defined as follows.

\smallskip{}

\begin{definition}\label{gamm_stab}The LPV system \eqref{ss_sys}
is $\gamma$-\emph{stabilizable} if a $\gamma<\infty$ and a function
$K_{2}\in\mathcal{F}\left(\gamma,P\right)$, where
\begin{equation}
\begin{array}{l}
\mathcal{F}\left(\gamma,P\right)\doteq\{f:\left\Vert f\left(0\right)\right\Vert _{\infty}<\infty,\\
\left\Vert f\left(p\right)-f\left(\widehat{p}\right)\right\Vert _{\infty}\leq\gamma\left\Vert p-\widehat{p}\right\Vert _{\infty},\forall p,\widehat{p}\in P\},
\end{array}\label{f0_ass}
\end{equation}
 exist such that the closed-loop system \eqref{eq:cl_sys} is input-to-state
$\ell_{\infty}$ stable.$\qquad\blacksquare$\end{definition}\smallskip{}

\begin{assumption} \label{ga_stab}The LPV system (\ref{ss_sys})
is $\gamma_{o}$-stabilizable, for some $\gamma_{o}<\infty$.$\qquad\blacksquare$\end{assumption}\smallskip{}

\begin{definition}\label{D:opt_inv}A function $K^{o}=(K_{1}^{o},K_{2}^{o})$
is an \emph{optimal controller} for the system (\ref{ss_sys}) if
\begin{equation}
K^{o}=\arg\min\limits _{K\in\mathscr{K}_{s}\cap\mathcal{F}\left(\gamma_{o},P\right)}GIE\left(K\right).\label{fo_def}
\end{equation}
where $\mathscr{K}_{s}$ is the set of all stabilizing controllers.$\qquad\blacksquare$\end{definition}
\smallskip{}

An optimal controller $K^{o}$ is thus a right-inverse function which,
among all functions stabilizing the closed-loop system, gives the
minimum global inversion error. Clearly, such a controller is not
known since, as assumed above, the system (\ref{ss_sys}) is not known
(even for known system, the evaluation of $K^{o}$ from (\ref{fo_def})
is in general hard).

In this paper, an approach is proposed, called \emph{Direct FeedbacK
(DFK) design}, that uses an approximation of $K^{o}$, identified
from the available data (\ref{D_inv}), as the controller $K$ in
the closed-loop system of Figure \ref{ctr_sys_1}. This approximation
is of the form
\begin{equation}
\begin{array}{c}
\widehat{K}=\left(\widehat{K}_{1},\widehat{K}_{2}\right)\\
\widehat{K}_{1}\left(p_{t}\right)=\left(\widehat{K}_{11}\left(p_{t}\right),\ldots,\widehat{K}_{1n_{x}}\left(p_{t}\right)\right)\\
\widehat{K}_{2}\left(p_{t}\right)=\left(\widehat{K}_{21}\left(p_{t}\right),\ldots,\widehat{K}_{2n_{x}}\left(p_{t}\right)\right)\\
\widehat{K}_{jl}\left(p_{t}\right)=\sum\limits _{i=1}^{m}a_{jli}\phi_{i}\left(p_{t}\right)
\end{array}\label{Kdef}
\end{equation}
where $\phi_{i}:P\rightarrow\mathbb{R}$ are Lipschitz continuous
basis functions and $a_{jli}\in\mathbb{R}$, $j=1,2$, $l=1,\ldots,n_{x}$,
$i=1,\ldots,m$. The vector $a=(a_{jli})\in\mathbb{R}^{N}$, with
$N\doteq2n_{x}m$, containing all the coefficients $a_{jli}$ is required
to be sparse, i.e. to have a ``small\textquotedblright{} number of
non-zero components. The main reason for this requirement is that
vector sparsity allows an efficient implementation on real-time processors,
which may have limited memory and computational capacity. Moreover,
sparse approximations are known to enjoy nice regularization properties,
providing good accuracy on new data and limiting well-known issues
such as over-fitting and the curse of dimensionality.

The following problem is investigated in this paper.\smallskip{}

\begin{problem}\label{des_prob}From the available data (\ref{D_inv}),
identify an approximation (\ref{Kdef}) of the optimal controller
$K^{o}$ such that:\\
(i) The feedback system of Fig. \ref{ctr_sys_1} is \emph{input-to-state}
$\ell_{\infty}$\emph{ stable}.\\
(ii) The vector $a=(a_{jli})\in\mathbb{R}^{N}$ is \emph{sparse}.
More precisely: the vector $a$ has a minimal $\ell_{1}$ quasi-norm.
\\
(iii) The\emph{ tracking error}
\[
TE_{t}\left(K\right)\doteq\left\Vert r_{t}-x_{t}\right\Vert _{\infty}
\]
is ``small\textquotedblright{} for all $t=0,1,\ldots\ $.$\qquad\blacksquare$
\end{problem}

\section{Closed-loop stability analysis}

\label{SS:DIC}

In this section, a theoretical result about stability of DFK control
systems is derived. This result will be used by the identification
Algorithm 1 in Section \ref{S:spid}, in order to obtain a controller
stabilizing the DFK closed-loop system.

Define the following \emph{residue function}:
\begin{equation}
\Delta\left(p_{t}\right)\doteq K^{o}\left(p_{t}\right)-\widehat{K}\left(p_{t}\right)\label{rf_def_1}
\end{equation}
where $K^{o}$ is an optimal controller (see Definition \ref{D:opt_inv})
and $\widehat{K}$ is an approximated controller. Considering that
$\widehat{K}$ and $K^{o}$ are Lipschitz continuous by definition,
we have that $\Delta$ is also Lipschitz continuous. Since $P$ is
a bounded set, it follows that
\begin{equation}
\lambda_{2}\doteq\max_{p\in P}\left\Vert \Delta_{2}\left(p\right)\right\Vert _{1}<\infty\label{delta_lip}
\end{equation}
where $\Delta_{2}\left(p_{t}\right)\doteq K_{2}^{o}\left(p_{t}\right)-\widehat{K}_{2}\left(p_{t}\right)$
and $\left\Vert \cdot\right\Vert _{1}$ is the $\ell_{1}$ vector
norm. Analogously, we have that
\begin{equation}
\lambda_{B}\doteq\max_{p\in P}\left\Vert B\left(p\right)\right\Vert _{\infty}<\infty\label{Bbound}
\end{equation}
where $\left\Vert \cdot\right\Vert _{\infty}$ is the $\ell_{\infty}$
vector norm. 

Consider now that the closed-loop system \eqref{eq:cl_sys} with $K=K^{o}$
is input-to-state $\ell_{\infty}$ stable by assumption. Thus, some
$\lambda_{S}<\infty$ and $\lambda_{e}<\infty$ exist such that the
state sequence $\mathbf{x}=(x_{1},x_{2},\ldots)$ is bounded as 
\[
\left\Vert \mathbf{x}\right\Vert _{\infty}\leq\lambda_{S}\left\Vert \mathbf{r}\right\Vert _{\infty}+\lambda_{e}\left\Vert \mathbf{e}\right\Vert _{\infty}
\]
for any reference sequence $\mathbf{u}=(u_{1},u_{2},\ldots)\in\ell_{\infty}$,
any noise sequence $\mathbf{e}=(e_{1},e_{2},\ldots)\in\ell_{\infty}$,
any parameter sequence $\mathbf{p}=(p_{1},p_{2},\ldots)\in\ell_{\infty}$
with $p_{t}\in P$, $\forall t$.

The following result gives sufficient conditions for the stability
of the DFK control scheme of Figure \ref{ctr_sys_1} with $K=\widehat{K}$
and provides a tight bound on the related tracking error. Tightness
is here intended in a worst-case sense: we say that a bound on a variable
is tight when the variable, in the worst possible case, hits the bound.
\smallskip{}

\begin{theorem}\label{thm_stab}Let Assumption \ref{ga_stab} hold.
If
\begin{equation}
\lambda_{2}<1/\lambda_{S},\label{stab_cond}
\end{equation}
then:\\
(i) The DFK closed-loop system \eqref{eq:cl_sys} with $K=\hat{K}$
is input-to-state $\ell_{\infty}$ stable.\\
(ii) The DFK closed-loop system tracking error is tightly bounded
as
\begin{equation}
TE_{t}\left(\widehat{K}\right)\leq IE\left(K^{o},w_{t-1},e_{t-1}\right)+\lambda_{B}\left\vert \Delta\left(w_{t-1}\right)\right\vert \label{teb}
\end{equation}
for all $t=0,1,\ldots$, where $w_{t}\doteq\left(p_{t},r_{t+1},x_{t}\right)$
and $\Delta\left(w_{t}\right)$ $\doteq$ $\Delta\left(p_{t}\right)(r_{t+1},-x_{t})$.\end{theorem}

\smallskip{}

\textbf{Proof.} The proof will be published on arXiv or on a journal
as soon as possible.$\qquad\blacksquare$

\section{DFK control design}

\label{S:spid}

\subsection{Available prior and experimental information}

\label{sub:info}

In this subsection, the information available for control design is
summarized.

\textbf{Experimental information. }The data (\ref{D_inv}) have been
collected, which can be conveniently described as 
\begin{equation}
\widetilde{u}_{k}=K^{o}\left(\widetilde{w}_{k}\right)+d_{k},\quad k=-L,\ldots,-1\label{Dset}
\end{equation}
where $K^{o}$ is the unknown optimal controller (see Definition \ref{D:opt_inv}),
$\widetilde{w}_{k}\doteq\left(\widetilde{p}_{k},\widetilde{x}_{k+1},\widetilde{x}_{k}\right)$,
$K^{o}\left(\widetilde{w}_{k}\right)\equiv K^{o}\left(\widetilde{p}_{k},\widetilde{x}_{k+1},\widetilde{x}_{k}\right)$
$\doteq$ $K^{o}\left(\widetilde{p}_{k}\right)(\widetilde{x}_{k+1},-\widetilde{x}_{k})$,
and $d_{k}$ accounts for the noise corrupting $u_{k}$ and for the
fact that $K^{o}$ is not an exact right-inverse of the function defining
the system (see the Appendix). Note that, in the data used for design,
the reference is replaced by the state at time $t+1$.$\qquad\blacksquare$

\textbf{Prior information on the noise }$d_{k}$\textbf{. }Since the
measurements $\widetilde{u}_{k}$ and $\widetilde{w}_{k}$ are bounded
and $K^{o}\in\mathcal{F}\left(\gamma_{o},P\right)$, where $K^{o}\equiv K^{o}\left(w_{t}\right)\doteq$
$K^{o}\left(p_{t}\right)(r_{t+1},-x_{t})$, $w_{t}\doteq\left(p_{t},r_{t+1},x_{t}\right)$,
it follows that the noise $d_{k}$ is bounded:
\begin{equation}
\left\vert d_{k}\right\vert \leq\delta,\quad k=-L,\ldots,-1\label{bnois}
\end{equation}
for some $\delta<\infty$.$\qquad\blacksquare$

\textbf{Prior information on the function }$\Delta$\textbf{.} The
function $\Delta\left(w_{t}\right)$ $\doteq$ $\Delta\left(p_{t}\right)(r_{t+1},-x_{t})$
is Lipschitz continuous:
\begin{equation}
\Delta\in\mathcal{F}\left(\gamma_{\Delta},\Omega_{\Delta}\right)\label{del_lip}
\end{equation}
for some $\gamma_{\Delta}<\infty$, where $\Omega_{\Delta}\doteq P\times X^{2}$.
This directly follows from (\ref{rf_def_1}) and from the Lipschitz
continuity of $K^{o}$ and $\widehat{K}$.$\qquad\blacksquare$

\subsection{DFK design algorithm}

\label{algo}

Based on the stability Theorem \ref{thm_stab}, an algorithm is proposed
in this subsection, for the direct design of LPV controllers allowing
us to satisfy the three requirements of Problem \ref{des_prob}.

Consider a set of Lipschitz continuous basis functions $\phi_{i}$,
$i=1,\ldots,m$. The choice of these functions can be carried out
considering the numerous options available in the literature (e.g.
Gaussian, sigmoidal, wavelet, polynomial, trigonometric, ...). Define
the matrix $\mathbf{\Psi}\in\mathbb{R}^{L\times N}$, $N\doteq2n_{x}m$,
as follows:
\begin{equation}
\mathbf{\Psi}\doteq\left[\begin{array}{cccccc}
-Z_{01}\Phi & \cdots & -Z_{0n_{x}}\Phi & Z_{11}\Phi & \cdots & Z_{1n_{x}}\Phi\end{array}\right]\label{psi_mat}
\end{equation}
where $Z_{ji}\in\mathbb{R}^{L\times L}$ and $\Phi\in\mathbb{R}^{L\times m}$
are given by 
\[
\begin{array}{l}
Z_{ji}\doteq\textrm{diag}\left(\left[\begin{array}{ccc}
\widetilde{x}_{i,j+1} & \cdots & \widetilde{x}_{i,j+L}\end{array}\right]\right)\\
\Phi\doteq\left[\begin{array}{ccc}
\phi_{1}\left(\widetilde{p}_{1}\right) & \cdots & \phi_{m}\left(\widetilde{p}_{1}\right)\\
\vdots & \ddots & \vdots\\
\phi_{1}\left(\widetilde{p}_{L}\right) & \cdots & \phi_{m}\left(\widetilde{p}_{L}\right)
\end{array}\right],
\end{array}
\]
$\widetilde{x}_{i,t}$ indicates the $i$th component of $\widetilde{x}_{t}$
and $j=0,1$. Define the sets of indices
\[
Q_{\zeta}^{k}\doteq\left\{ i:\left\Vert \left(\widetilde{p}_{k},\widetilde{x}_{k+1}\right)-\left(\widetilde{p}_{i},\widetilde{x}_{i+1}\right)\right\Vert _{\infty}\leq\zeta\right\} 
\]
where $k=-L,\ldots,-1$ and $\zeta$ is the minimum value for which
every index set $Q_{\zeta}^{k}$ contains at least two elements.\smallskip{}

\textbf{Algorithm 1}
\begin{enumerate}
\item \label{est_delta}Estimate $\delta$ and $\lambda_{S}$ by means of
the validation procedure in \cite{MiNoAUT04} from the data set (\ref{D_inv}).

\item \label{st_gg}Choose $\lambda_{2}^{s}<1/\lambda_{S}$.
\item \label{st_1}Solve the optimization problem 
\begin{equation}
\begin{array}{l}
b^{*}=\arg\min\limits _{b\in\mathbb{R}^{N}}\left\Vert b\right\Vert _{1}\\
\text{subject to}\\
(a)\ \left\Vert \widetilde{\mathbf{u}}-\mathbf{\Psi}b\right\Vert _{\infty}\leq\delta\\
(b)\ \left\vert \widetilde{u}_{l}-\widetilde{u}_{k}+\left(\mathbf{\Psi}_{k}-\mathbf{\Psi}_{l}\right)b\right\vert \leq\lambda_{2}^{s}\left\Vert \widetilde{x}_{l}-\widetilde{x}_{k}\right\Vert _{\infty}+2\delta,\\
\qquad\qquad\qquad\qquad\qquad k=-L,\ldots,-1,\ l\in Q_{\zeta}^{k}
\end{array}\label{opt21a}
\end{equation}
where $\widetilde{\mathbf{u}}\doteq(\widetilde{u}_{-L},\ldots,\widetilde{u}_{-1})$
and $\mathbf{\Psi}_{k}$ denotes the $k$th row of the matrix $\mathbf{\Psi}$.$\qquad\blacksquare$
\end{enumerate}
The DFK controller is 
\begin{equation}
K^{\ast}\left(p_{t}\right)\doteq(K_{1}^{\ast}\left(p_{t}\right),K_{2}^{\ast}\left(p_{t}\right))\label{eq:kstar}
\end{equation}
where
\[
\begin{array}{c}
K_{1}^{\ast}\left(p_{t}\right)=(K_{11}^{\ast}\left(p_{t}\right),\ldots,K_{1n_{x}}^{\ast}\left(p_{t}\right))\\
K_{2}^{\ast}\left(p_{t}\right)=(K_{21}^{\ast}\left(p_{t}\right),\ldots,K_{2n_{x}}^{\ast}\left(p_{t}\right))
\end{array}
\]
\begin{equation}
K_{jl}^{\ast}\left(p_{t}\right)=\sum_{i=1}^{m}a_{jli}^{\ast}\phi_{i}\left(p_{t}\right)\label{f_star}
\end{equation}
and $a_{jli}^{\ast}=b_{(j-1)n_{x}m+(l-1)m+i}^{\ast}$.$\qquad\blacksquare$

\subsection{Asymptotic stability property}

\label{stab_prop}

A result is now presented, showing that the controller $K^{\ast}$
identified by Algorithm $1$ satisfies the stability condition (\ref{stab_cond})
when the number of data $L$ tends to infinity. The following assumption
is needed to prove this result.\smallskip{}

\begin{assumption} \label{dense_ass} The set of points $\mathcal{D}_{wd}^{L}\doteq\left\{ \left(\widetilde{w}_{k},d_{k}\right)\right\} _{k=-L}^{-1}$
is dense on $\Omega_{\Delta}\times\mathcal{B}_{\delta}$ as $L\rightarrow\infty$,
where $\mathcal{B}_{\delta}\doteq\left\{ d\in\mathbb{R}:\left\vert d\right\vert \leq\delta\right\} $.
That is, for any pair $(w,d)\in\Omega_{\Delta}\times\mathcal{B}_{\delta}$
and any $\lambda>0$, a $L_{\lambda}<\infty$ and a pair $\left(\widetilde{w}_{k},d_{k}\right)\in\mathcal{D}_{wd}^{L_{\lambda}}$
exist such that $\left\Vert (y,d)-\left(\widetilde{w}_{k},d_{k}\right)\right\Vert _{\infty}\leq\lambda.$$\qquad\blacksquare$

\end{assumption}\smallskip{}

Assumption \ref{dense_ass} essentially ensures that the controller
domain $\Omega_{\Delta}$ is ``well explored\textquotedblright{}
by the data $\widetilde{w}_{k}$ and, at the same time, the noise
$d_{k}$ covers its domain $\mathcal{B}_{\delta}$, hitting the bounds
$-\delta$ and $\delta$ with arbitrary closeness after a sufficiently
long time. This latter noise property is called tightness, see \cite{NoFaMiAUT13}
and, for a probabilistic version, \cite{BCT98}.\smallskip{}

\begin{theorem}\label{lip_conv}Let the optimization problem (\ref{opt21a})
be feasible for any $L>0$. Let Assumption \ref{dense_ass} holds.
Then, $\max_{p\in P}\left\Vert \Delta_{2}\left(p\right)\right\Vert _{1}=\lambda_{2}$,
where
\begin{equation}
\limsup_{L\rightarrow\infty}\lambda_{2}\leq\lambda_{2}^{s}<1/\lambda_{S}.\label{la2_lim}
\end{equation}
\end{theorem}\smallskip{}

\textbf{Proof.} The proof will be published on arXiv or on a journal
as soon as possible.$\qquad\blacksquare$

\subsection{Set Membership optimality analysis}

\label{sec:opt_an}

From the information summarized in Subsection \ref{sub:info}, we
have that $K^{o}\in FIFS$, where $FIFS$ is the \emph{Feasible Inverse
Function Set}, defined as 
\[
\begin{array}{r}
FIFS\doteq\{K=\widehat{K}+\Delta:\Delta\in\mathcal{F}\left(\gamma_{\Delta},\Omega_{\Delta}\right),\\
\left\vert \widetilde{u}_{k}-K\left(\widetilde{w}_{k}\right)\right\vert \leq\delta,k=-L,\ldots,-1\}.
\end{array}
\]
According to this definition, $FIFS$ is the set of all inverse functions
consistent with the prior and experimental information. Hence, the
tightest bound on $\left\vert \Delta\left(w_{t-1}\right)\right\vert $
$=$ $|K^{o}\left(w_{t-1}\right)$ $-$ $\widehat{K}\left(w_{t-1}\right)|$
which can be derived on the basis of this information is given by
$\sup\limits _{K\in FIFS}|K^{o}\left(w_{t-1}\right)-\widehat{K}\left(w_{t-1}\right)|,$
leading to the following definition of \emph{worst-case tracking error}:
\begin{equation}
WE_{t}\left(\widehat{K}\right)\doteq IE\left(K^{o},w_{t-1},e_{t-1}\right)+\lambda_{B}AE_{t}\left(\widehat{K}\right)\label{eq:we_def}
\end{equation}
where 
\[
AE_{t}\left(\widehat{K}\right)\doteq\sup\limits _{K\in FIFS}\left\vert K\left(w_{t-1}\right)-\widehat{K}\left(w_{t-1}\right)\right\vert 
\]
is called the \emph{worst-case approximation error}. An optimal DFK
controller is defined as an inverse function $K^{op}$ which guarantees
the closed-loop stability and minimizes the worst-case approximation
error. However, finding an optimal DFK controller may be hard or not
convenient from a computational point of view, and sub-optimal solutions
can be looked for. In particular, approximations that guarantee a
degradation of at most 2 are often considered in the literature. These
approximations are called almost-optimal \cite{Traub88}, \cite{MiNorLaWa96}.\smallskip{}

\begin{definition}An inverse function $K^{ao}$ is an \emph{almost-optimal
DFK controller\ }if:\\
(i) $\max_{p_{t}\in P}\left\Vert K_{2}^{o}\left(p_{t}\right)-K_{2}^{ao}\left(p_{t}\right)\right\Vert _{1}<1/\lambda_{S}$.\\
(ii) $AE_{t}\left(K^{ao}\right)\leq2\inf_{K}AE_{t}\left(K\right)$,
for all $t=0,1,\ldots$ .$\qquad\blacksquare$\end{definition}

\smallskip{}

The following theorem shows that, provided the closed-loop stability
(which can be guaranteed when the number of data is large, see Theorem
\ref{lip_conv}), the controller $K^{\ast}$ identified by means of
Algorithm $1$ is almost-optimal.\smallskip{}

\begin{theorem}\label{nsm_opt}Let the optimization problem (\ref{opt21a})
be feasible. Let Assumption \ref{ga_stab} holds. Assume also that
\begin{equation}
\max_{p\in P}\left\Vert K_{2}^{o}\left(p_{t}\right)-K_{2}^{ao}\left(p_{t}\right)\right\Vert _{1}<1/\lambda_{S}\label{stab_ass2}
\end{equation}
Then, the DFK controller $K^{\ast}$ is almost-optimal.\end{theorem}\smallskip{}

\textbf{Proof.} See the Appendix.$\qquad\blacksquare$

\section{Example 1: control of the Duffing oscillator}

\label{example_duffing}

The Duffing system is a second-order damped oscillator with nonlinear
spring, described by the following differential equations:
\begin{equation}
\begin{array}{l}
\dot{x}_{1}=x_{2}\\
\dot{x}_{2}=-\alpha_{1}x_{1}-\alpha_{2}x_{1}^{3}-\beta x_{2}+u
\end{array}\label{duffing}
\end{equation}
where $x=(x_{1},x_{2})$ is the system state ($x_{1}$ and $x_{2}$
are the oscillator position and velocity, respectively) and $u$ is
the input. The following values of the parameters have been considered:
$\alpha_{1}=1$, $\alpha_{2}=-1$, $\beta=0.2$. For these parameter
values and for certain choices of the input signal, this system exhibits
a chaotic behavior, and this makes control design a particularly challenging
problem. Note that this is a nonlinear system which can be seen as
a quasi-LPV system, according to the definition given in \cite{ShRu00}.

A simulation of the Duffing system (\ref{duffing}) having duration
$200$ s has been performed, using the input signal $u(\tau)=0.4\sin(\tau)$
($\tau$ here denotes the continuous time). A set of $L=2000$ data
have been collected from this simulation with a sampling period $T_{s}=0.1$
s:
\[
\mathcal{D}\doteq\left\{ \widetilde{p}_{k},\widetilde{x}_{k},\widetilde{u}_{k}\right\} _{k=-2000}^{-1}
\]
where $\widetilde{u}_{k}$ are the measurements of the input, $\widetilde{x}_{k}$
are the measurements of the state, corrupted by a zero-mean Gaussian
noise having a noise-to-signal standard deviation ratio of $0.05$,
and the scheduling parameter is $\widetilde{p}_{t}=\widetilde{x}_{t}$.

A set of $m=28$ polynomial basis functions has been considered:
\[
\begin{array}{c}
\phi_{1}\left(p\right)=1,\phi_{2}\left(p\right)=p_{1},\phi_{3}\left(p\right)=p_{2},\\
\phi_{4}\left(p\right)=p_{1}^{2},\phi_{5}\left(p\right)=p_{1}p_{2},\phi_{6}\left(p\right)=p_{2}^{2},\\
\ldots,\phi_{27}\left(p\right)=p_{1}p_{2}^{5},\phi_{28}\left(p\right)=p_{2}^{6}.
\end{array}
\]
The matrix $\mathbf{\Psi}$ has been constructed according to (\ref{psi_mat}).
Note that the total number of basis functions used in the matrix $\mathbf{\Psi}$
is $N=4m=112$. Larger polynomial degrees have been considered, giving
larger set of basis functions, but no significant improvements have
in general been observed.

Then, Algorithm 1 has been applied: The bound $\delta$ in (\ref{bnois})
has been estimated (together with the Lipschitz constant $\gamma_{\Delta}$)
from the data set $\mathcal{D}$ by means of the validation procedure
in \cite{MiNoAUT04}, suitably adapted for the present LPV case. The
bound $\lambda_{B}$ in (\ref{stab_cond}) has also been estimated
from the data set $\mathcal{D}$, using a similar adaption of the
procedure in \cite{MiNoAUT04}. The optimization problem in Algorithm
1 has been solved using the CVX toolbox \cite{cvx}. A controller
$K^{\ast,1}=(K_{1}^{\ast,1},K_{2}^{\ast,1})$ of the form (\ref{f_star})
has been obtained. 

In order to verify the generalization capability of the controller,
a validation test has been carried out. A new set of data has been
generated by performing a simulation of the Duffing system (\ref{duffing})
with duration $200$ s (corresponding to $2000$ data). In this simulation,
an input signal different from the one of the design data set has
been considered. In particular, the input used for the validation
data is $u(\tau)=0.3\sin(0.5\tau)+\xi_{G}(\tau)$, where $\xi_{G}(\tau)$
is a white Gaussian noise with zero mean and standard deviation $0.3$.
The controller has been simulated on both the design and validation
data and, for each of these two simulations, the $RMS\left(\widetilde{u}_{t}-\hat{u}_{t}\right)$
index has been computed, where $\widetilde{u}_{t}$ is the measured
input, $\hat{u}_{t}$ is the input provided by the controller and
$RMS\left(z_{t}\right)\doteq\sqrt{\frac{1}{2000}\sum\nolimits _{t=0}^{1999}z_{t}^{2}}.$
The $RMS$ values $0.0903$ and $0.1094$ have been obtained on the
design and validation set, respectively. The $RMS$ value obtained
on the validation set is thus very similar to that obtained on the
design set, indicating that the controller is able to provide the
correct command input also for trajectories different from those used
for its design. Obviously, this does not guarantee closed-loop stability.

Note that, as for any control design method, verifying theoretically
the stabilizability of a real system is not easy in general (or even
not possible). What can be done in practice is checking this assumption
on the measured data, first estimating $\lambda_{S}$ by means of
the validation procedure in \cite{MiNoAUT04}, then checking if Lipschitz
continuous basis functions and a $\lambda_{2}^{s}<1/\lambda_{S}$
exist for which the optimization problem of Algorithm 1 is feasible.
This numerical verification is quite easy to carry out, since the
validation procedure consists in applying a simple data pre-processing
algorithm and the feasibility check can be performed just verifying
consistency with the data, \cite{MiNoAUT04}. The Duffing system considered
here resulted to be stabilizable.

\begin{figure}[h]
\begin{centering}
\includegraphics[scale=0.5]{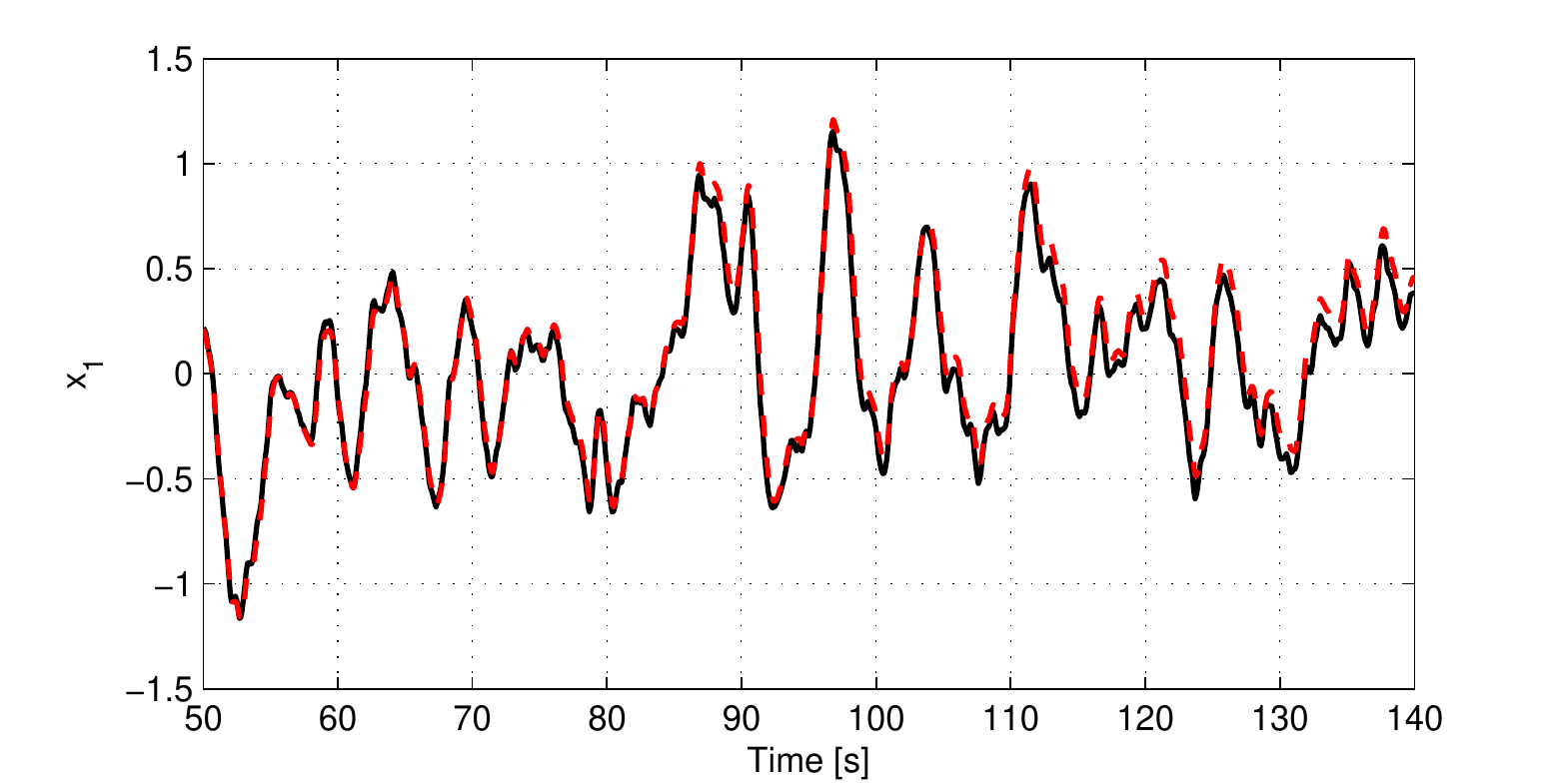} 
\par\end{centering}

\begin{centering}
\includegraphics[scale=0.5]{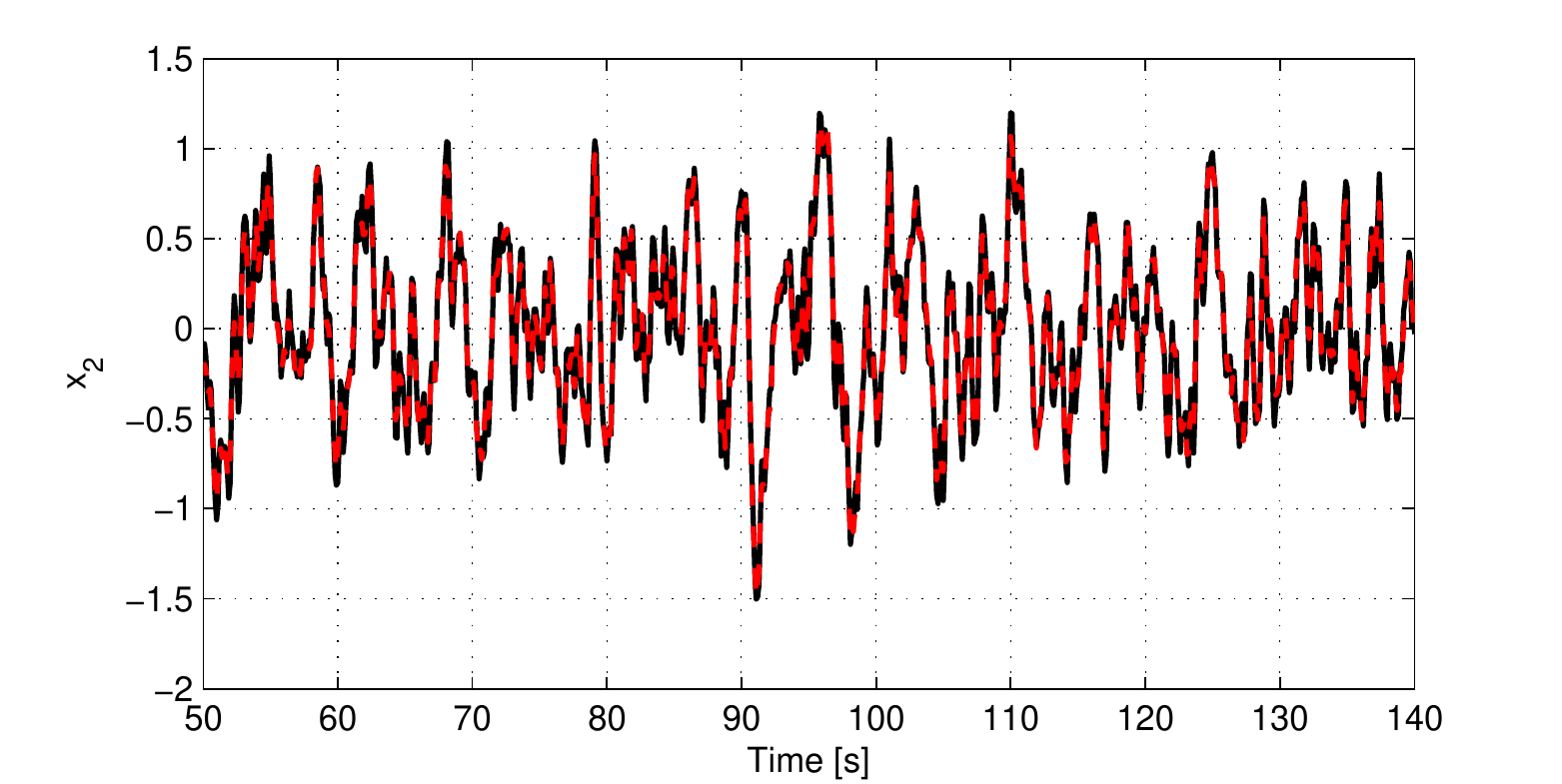} 
\par\end{centering}

\protect\caption{Duffing oscillator. Continuous (black) line: reference. Dashed (red)
line: actual state.}

\label{comp} 
\end{figure}

\begin{figure}[h]
\begin{centering}
\includegraphics[scale=0.5]{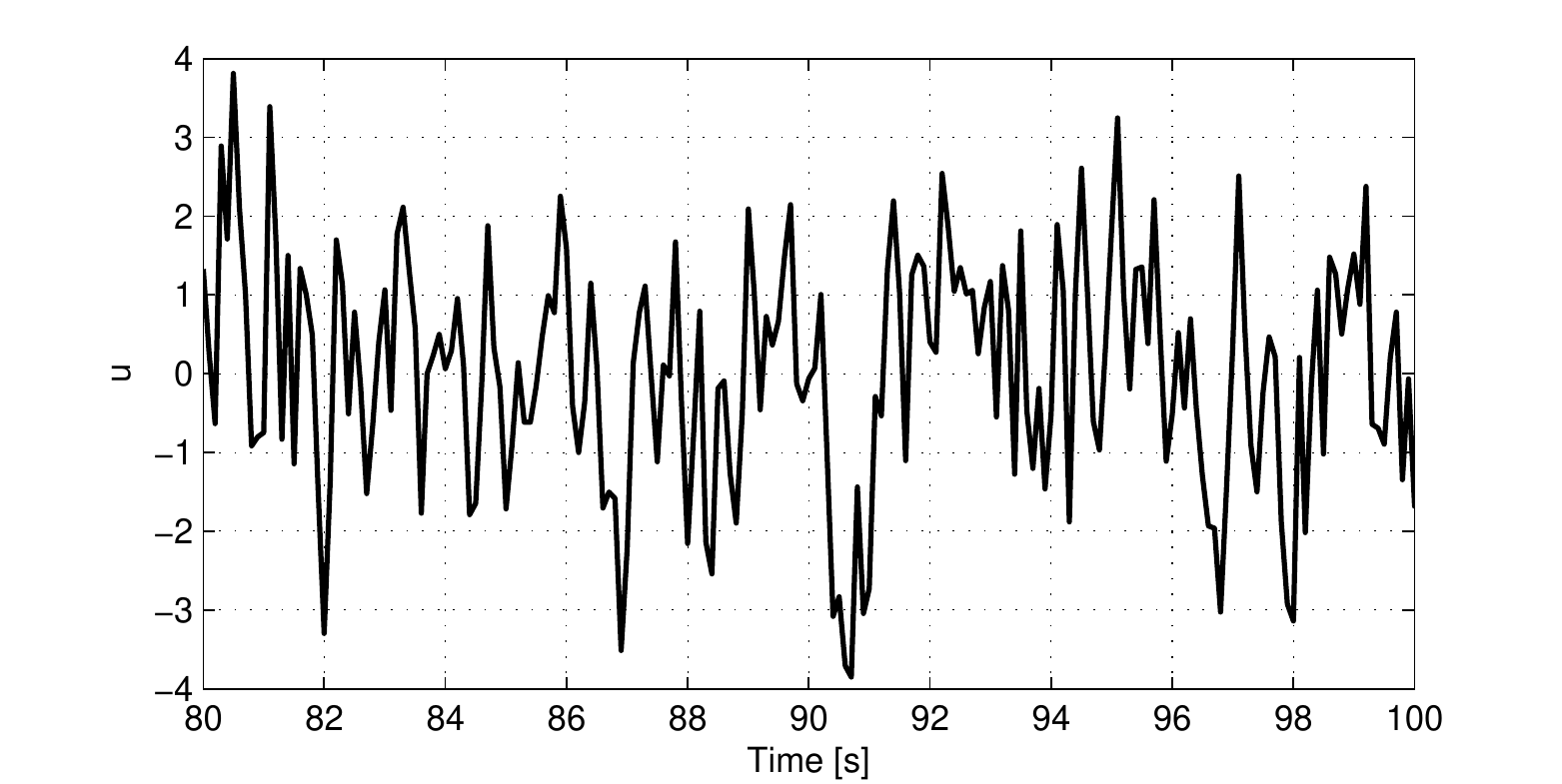} 
\par\end{centering}

\protect\caption{Duffing oscillator. Command input corresponding to Figure \ref{comp}.}

\label{fig:ut} 
\end{figure}

The DFK control scheme of Figure \ref{ctr_sys_1} has then been implemented,
where $S$ is the system (\ref{duffing}) (including D/A and A/D converters,
the former preceding, the latter following the system), $K_{1}=K_{1}^{\ast,1}$,
$K_{2}=K_{2}^{\ast,1}$, and $e_{t}$ is a Gaussian noise affecting
the state measurements, having zero-mean and a noise-to-signal standard
deviation ratio of $0.05$. A simulation of the DFK control system
with duration $200$ s has been performed, using zero initial conditions
and a reference signal $r_{t}$ with first component (the oscillator
position) generated as a uniform noise with amplitude $5$, filtered
by a second-order filter with a cutoff frequency of $1$ rad/s (this
filter has been inserted in order to ensure not too high variations).
The second component (the oscillator velocity) has been generated
as the derivative of the first. The state variables have been corrupted
by zero-mean Gaussian noises with a noise-to-signal standard deviation
ratio of $0.05$. In Figure \ref{comp}, the two states are compared
to their respective references for a portion of the simulation time
interval. The corresponding command input $u_{t}$ is shown in Figure
\ref{fig:ut} for a smaller portion of the simulation time interval.

A second DFK controller, called $K^{\ast,2}$, has been designed using
the scheduling variable $\widetilde{p}_{t}=\widetilde{x}_{1,t}^{2}$,
where $\widetilde{x}_{1,t}$ is the first component of the measured
state vector $\widetilde{x}_{t}$. Indeed, with this choice, the Duffing
system (\ref{duffing}) becomes a ``true'' LPV system. This design
has been performed by means of Algorithm 1, using polynomial basis
functions (up to the $6^{th}$ degree) and the same data set and procedure
employed for the design of the first controller.

A third DFK controller, called $K^{\ast,3}$, has been also designed,
assuming that the full state is not available but only the output
$\tilde{y}_{t}=\widetilde{x}_{1,t}$ can be measured. Feedback has
been performed from the regressor $\left(\tilde{y}_{t},\tilde{y}_{t-1}\right)$
and the scheduling variable has been chosen as $\widetilde{p}_{t}=\left(\tilde{y}_{t},\tilde{y}_{t-1}\right)$.
This design has been carried out by means of Algorithm 1, using polynomial
basis functions (up to the $6^{th}$ degree) and the same data set
and procedure employed for the design of the other two controllers.
For $K^{\ast,3}$, the second component of the reference has been
taken as the delayed first component. Note that, for $K^{\ast,3}$,
tracking regards only the output $\tilde{y}_{t}$, i.e. the first
state component.

Then, a Monte Carlo (MC) simulation has been carried out, where this
data-generation-control-design procedure has been repeated 100 times.
For each trial, the tracking performance has been evaluated by means
of the Root Mean Square tracking errors $RMS_{i}=RMS\left(r_{i,t}-x_{i,t}\right)$,
where $i$ indicates the $i$th vector component and 
\begin{equation}
RMS\left(z_{t}\right)\doteq\sqrt{\frac{1}{2000}\sum\nolimits _{t=0}^{1999}z_{t}^{2}}.\label{eq:rms}
\end{equation}
 The averages $\overline{RMS}_{i}$ of these errors obtained in the
MC simulation are reported in Table \ref{tab1}, together with the
average number $\bar{N}^{sel}$ of basis functions selected by Algorithm
1. From these results, it can be concluded that all the DFK controllers
are able to ensure an accurate tracking, even in the presence of quite
high measurement noises, using a small number of basis functions.

In order to study the behavior of a DFK control system in function
of the used basis functions, further trials have been performed, relaxing
the constrains in \eqref{opt21a} (i.e. increasing $\delta$). The
setting used for the controller $K^{\ast,1}$ has been considered
in these trials. The obtained results can be summarized as follows:
with 21 basis functions, a deterioration of the $RMS$ tracking errors
of about 2 times has been observed; with 17 basis functions, a deterioration
of the $RMS$ tracking errors of about 5 times has been observed;
with 5 basis functions, the controller was still able to stabilize
the closed-loop system but the actual state was completely different
from the reference.

Note that the control design procedure is very simple (in all the
three cases), not requiring to identify an LTI model for each working
condition as in gain-scheduling or to solve an optimization problem
at each time step as in predictive control. In the DFK approach, the
controller is designed just solving off-line a convex optimization
problem, using directly the measured data. The resulting controllers
are also extremely simple: only a few basis functions are used to
control the system over its whole domain. Moreover, the DFK approach
is intrinsically robust, since it is based on the direct design of
the controller from experimental data, avoiding thus any possible
model error.

\begin{table}
\centering

\begin{tabular}{|c||c|c|c|}
\hline 
 & $K^{\ast,1}$ & $K^{\ast,2}$ & $K^{\ast,3}$\tabularnewline
\hline 
\hline 
$\bar{N}^{sel}$ & 27 & 26 & 27\tabularnewline
\hline 
$\overline{RMS}_{1}$ & 0.0562 & 0.0723 & 0.0534\tabularnewline
\hline 
$\overline{RMS}_{2}$ & 0.0859 & 0.1792 & -\tabularnewline
\hline 
\end{tabular}\smallskip{}

\protect\caption{Duffing oscillator. Average number of selected basis function and
average $RMS$ tracking errors.}

\label{tab1}
\end{table}

\section{Example 2: Control of a 2-DOF robot manipulator}

\label{example2}

\begin{figure}[h]
\begin{centering}
\includegraphics[scale=0.7]{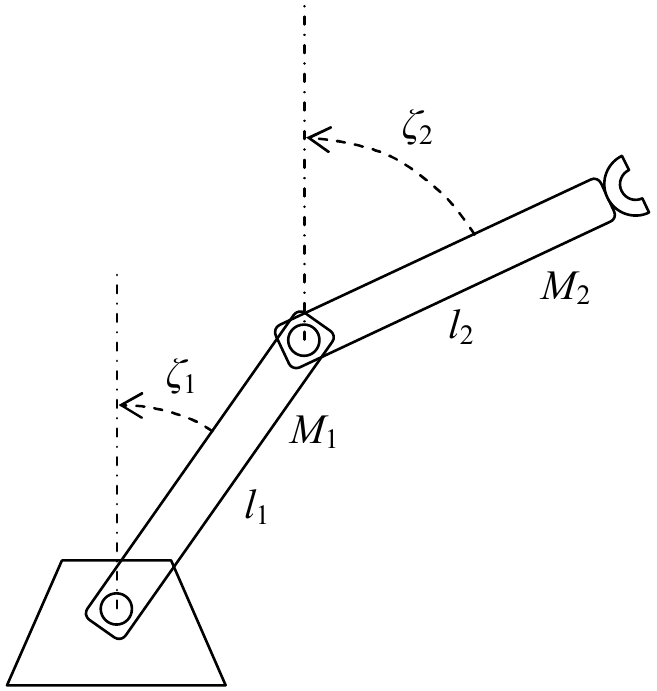} 
\par\end{centering}

\protect\caption{Robot Manipulator. }

\label{fig20} 
\end{figure}

The 2-DOF (2-degrees of freedom) robot manipulator depicted in Figure
\ref{fig20} has been considered in this example. In this Figure,
$\zeta_{1}$ and $\zeta_{2}$ are the angular positions of the two
segments of the robot arm, $u_{1}$ and $u_{2}$ are the control torques
acting on these segments, $l_{1}$ and $l_{2}$ are the segment lengths,
and $M_{1}$ and $M_{2}$ are the segment masses. This robot manipulator
can be described by the following continuous-time state-space LPV
model:
\begin{equation}
\dot{z}(\tau)=A^{c}(p(\tau))z(\tau)+B^{c}(p(\tau))u(\tau)\label{robot_ss}
\end{equation}
where $\tau$ is the continuous time, $z(\tau)=[\zeta_{1}(\tau)$
$\zeta_{2}(\tau)$ $\dot{\zeta}_{1}(\tau)$ $\dot{\zeta}_{2}]^{\top}$
is the state, $u(\tau)=[u_{1}(\tau)$ $u_{2}(\tau)]^{\top}$ is the
input, $p(\tau)=z(\tau)$, $A^{c}(p(\tau))\in\mathbb{R}^{4\times4}$,
$B^{c}(p(\tau))\in\mathbb{R}^{4\times2}$, $C=[I$ $\mathbf{0}]$,
$I$ is the $2\times2$ identity matrix, and $\mathbf{0}$ is the
$2\times2$ zero matrix. The expression of matrices $A^{c}(p(\tau))$
and $B^{c}(p(\tau))$ can be found in \cite{KwWe05}.

A set of $L=5000$ data has been generated by simulation of the system
(\ref{robot_ss}), with $l_{1}=0.8$ m, $l_{2}=0.7$ m, $M_{1}=2.5$
Kg, $M_{2}=2$ Kg: 
\[
\mathcal{D}\doteq\left\{ \widetilde{p}_{k},\widetilde{x}_{k},\widetilde{u}_{k}\right\} _{k=-5000}^{-1}.
\]
The data have been collected with a sampling time $T_{s}=0.02$ s,
using the following input signals:
\begin{equation}
u_{j}(\tau)=\left\{ \begin{array}{l}
-20z_{j}(\tau),\text{ if }\left\vert z_{j}(\tau)\right\vert \geq1.75\text{ rad}\\
0,\text{\quad if }l<\tau\leq l+500,\text{ }l=500,1500,2500,3500,\\
\qquad\text{and }\left\vert z_{j}(\tau)\right\vert <1.75\\
U\sin(\omega_{j1}\tau)+U\sin(\omega_{j2}\tau),\text{\quad otherwise,}
\end{array}\right.\label{u_rob}
\end{equation}
where $U=100$ Nm, $\omega_{11}=0.07$ rad/s, $\omega_{12}=0.8$ rad/s
$\omega_{21}=0.08$ rad/s $\omega_{11}=0.9$ rad/s, and $j=1,2$.
The feedback input on the first line of (\ref{u_rob}) has been applied
in order to limit the working range of $z_{1}$ and $z_{2}$ to the
interval $[-\pi,\pi]$ rad (the gain $-20$ and the threshold $1.75$
rad have been chosen thorough several preliminary simulations). Measurement
noises have been added to $z_{j}$, $j=1,\ldots,4,$ simulated as
uniform noises with amplitude $0.01$ rad. 

From these data, a DFK controller with two outputs has been designed.
In particular, for each component of $u$, a controller of the form
\ref{eq:kstar} has been designed using Algorithm 1. The two single
output controllers have been combined to obtain the two output overall
controller. For both the single output controllers, polynomial basis
functions up to degree $6$ have been considered.

\begin{figure}[h]
\begin{centering}
\includegraphics[scale=0.6]{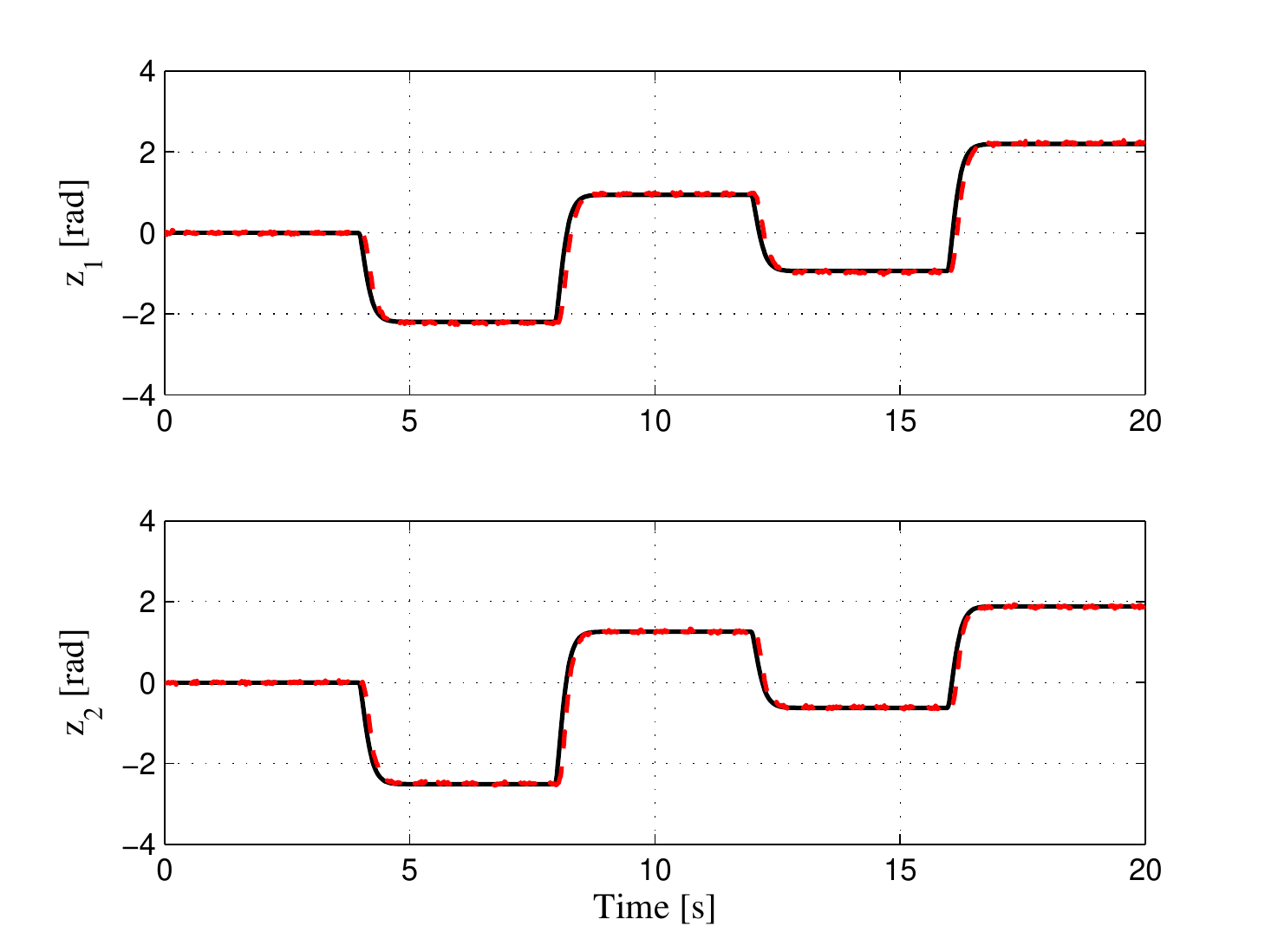} 
\par\end{centering}

\protect\caption{Robot Manipulator. Continuous (black) line: reference. Dashed (red)
line: actual state.}

\label{fig22} 
\end{figure}

Then, a simulation has been performed to test the DFK controller in
the task of reference tracking. Zero initial conditions have been
assumed. A reference signal of length $5000$ samples (corresponding
to $100$ s) has been used, defined as a sequence of step signals
with amplitudes in the interval $[-\pi,\pi]$, filtered by a second-order
filter with a cutoff frequency of $10$ rad/s (this filter has been
inserted in order to ensure not too high variations). The state variables
have been corrupted by uniform noises with amplitude $0.01$ rad.
In Figure \ref{fig22}, the angular positions of the controlled system
are compared with the references for the first $20$ s of this simulation.
Note that the two position references have been chosen quite similar
to each other (but not equal) in order to allow the manipulator to
reach in the most simple way any position in its range. 

The DFK control system has also been compared with the one in \cite{ChLPV11},
designed by means of a two-step method, involving LPV model identification
and Gain Scheduling (GS) control. The $RMS$ tracking errors (see
\eqref{eq:rms}) obtained in the simulation of the two control systems
for the two angular positions are reported in Table \ref{tab1-1}.

From these results, it can be observed that the control system is
quite effective, showing a fast tracking and a satisfactory steady-state
precision. In comparison with the method of \cite{ChLPV11}, DFK is
significantly simpler, both in the design phase (no models are required)
and in the implementation (the DFK controller is defined by a simple
function, involving ``a few'' basis functions).

\begin{table}
\centering

\begin{tabular}{|c||c|c|}
\hline 
 & DFK & GS\tabularnewline
\hline 
\hline 
$RMS_{1}$ & 0.172 & 0.167\tabularnewline
\hline 
$RMS_{2}$ & 0.148 & 0.152\tabularnewline
\hline 
\end{tabular}\smallskip{}

\protect\caption{Robot Manipulator. $RMS$ tracking errors.}

\label{tab1-1}
\end{table}

\section{Appendix}

\textbf{Noise in \eqref{ss_sys}.} Consider an LPV system in state-space
form, including also the measurement equation:
\[
\begin{array}{l}
z_{t+1}=A\left(p_{t}\right)z_{t}+B\left(p_{t}\right)u_{t}+Q\left(p_{t}\right)\xi_{t}^{p}\\
x_{t}=z_{t}+\xi_{t}^{m}
\end{array}
\]
where $z_{t}$ is the state, $x_{t}$ is the measured state, $u_{t}$
is the input, $p_{t}$ is the scheduling parameter vector, $\xi_{t}^{p}$
is a process disturbance and $\xi_{t}^{m}$ is a measurement noise.
Taking $z_{t+1}$ from the first equation and replacing it in the
second one (with $t\rightarrow t+1$), we have
\[
\begin{array}{l}
x_{t+1}=A\left(p_{t}\right)z_{t}+B\left(p_{t}\right)u_{t}+Q\left(p_{t}\right)\xi_{t}^{p}+\xi_{t+1}^{m}\\
=A\left(p_{t}\right)x_{t}+B\left(p_{t}\right)u_{t}+Q\left(p_{t}\right)\xi_{t}^{p}+\xi_{t+1}^{m}-A\left(p_{t}\right)\xi_{t}^{m}.
\end{array}
\]
Posing 
\[
\begin{array}{l}
H(p_{t})=[Q\left(p_{t}\right),I,-A\left(p_{t}\right)]\\
e_{t}=\left[\begin{array}{c}
\xi_{t}^{p}\\
\xi_{t+1}^{m}\\
\xi_{t}^{m}
\end{array}\right],
\end{array}
\]
we obtain the state equation
\[
x_{t+1}=A\left(p_{t}\right)x_{t}+B\left(p_{t}\right)u_{t}+H\left(p_{t}\right)e_{t}
\]
where $e_{t}$ includes both process and measurement disturbances.$\qquad\blacksquare$

\textbf{Noise in \eqref{Dset}.} According to the state equation,
$u_{t}$ is the input value that, for given $p_{t}$, $x_{t}$ and
$e_{t}$, yields the value $x_{t+1}$ of the state at time $t+1$.
On the other hand, the optimal controller $K^{o}$ provides, for $r_{t+1}=x_{t+1}$,
an input value $u_{t}^{o}$ given by
\[
u_{t}^{o}=K^{o}\left(p_{t},x_{t+1},x_{t}\right).
\]
If $K^{o}$ is an exact right-inverse of $g$, then $u_{t}=u_{t}^{o}$.
Otherwise, $u_{t}=u_{t}^{o}+\xi_{t}^{o}$ where $\xi_{t}^{o}$ is
an error due to the fact that $K^{o}$ is not an exact right-inverse
of the function defining the LPV system.

Suppose that a set of noise-corrupted measurements is available:
\[
\mathcal{D}\doteq\left\{ \widetilde{p}_{k},\widetilde{x}_{k},\widetilde{u}_{k}\right\} _{k=1}^{L}
\]
where $\tilde{u}_{t}=u_{t}+\xi_{t}^{u}$ is the measured input, $\xi_{t}^{u}$
is a measurement noise, $\widetilde{x}_{k}=x_{k}$ (note that $x_{k}$
is already the measured state, including process and measurement noises)
and $\widetilde{p}_{k}=p_{k}$ is the measured scheduling vector (for
simplicity but without loss of generality, $p_{k}$ is assumed to
be noise-free). Then, we can write the above equation (i.e. $u_{t}^{o}=K^{o}\left(p_{t},x_{t+1},x_{t}\right)$)
as
\[
\tilde{u}_{t}=K^{o}\left(\tilde{p}_{t},\tilde{x}_{t+1},\tilde{x}_{t}\right)+\xi_{t}^{o}+\xi_{t}^{u}=K^{o}\left(\tilde{w}_{t}\right)+d_{t}
\]
where $\tilde{w}_{t}=\left(\tilde{p}_{t},\tilde{x}_{t+1},\tilde{x}_{t}\right)$
and $d_{t}=\xi_{t}^{o}+\xi_{t}^{u}$. From this equation, we can see
that $d_{t}$ accounts for the noise corrupting $u_{t}$ and for the
fact that $K^{o}$ is not an exact right-inverse of the function defining
the LPV system. The noise $e_{t}$ is implicitly taken into account
in the argument of the controller.$\qquad\blacksquare$

\textbf{Proof of Theorem \ref{nsm_opt}. }(i) If the optimization
problem (\ref{opt21a}) is feasible, then the function $K^{\ast}$
defined in (\ref{f_star}) exists. Now consider that $K^{\ast}=K^{\ast}+\Delta$,
with $\Delta=0$. Obviously, if $\Delta=0$, then $\Delta\in\mathcal{F}\left(\gamma_{\Delta},\Omega_{\Delta}\right)$,
for any $\gamma_{\Delta}\geq0$. Moreover, from (\ref{opt21a}), $\left\Vert \widetilde{\mathbf{u}}-K^{\ast}\left(\widetilde{\mathbf{w}}\right)\right\Vert _{\infty}\leq\delta$,
implying that $K^{\ast}\in FIFS$. This guarantees that 
\[
\begin{array}{c}
AE_{t}\left(K^{\ast}\right)=\sup\limits _{K\in FIFS}\left\vert K\left(w_{t-1}\right)-K^{\ast}\left(w_{t-1}\right)\right\vert \\
\leq2\inf\limits _{\widehat{K}}\sup\limits _{K\in FIFS}\left\vert K\left(r_{t},x_{t-1}\right)-\widehat{K}\left(r_{t},x_{t-1}\right)\right\vert 
\end{array}
\]
see \cite{Traub88}, \cite{MiNorLaWa96}. Then, the claim follows
from the stability condition (\ref{stab_cond}) and the definition
of almost-optimal DFK controller.$\qquad\blacksquare$

\bibliographystyle{plain}
\bibliography{lettaltr,lettnos,lpv,sparsification,lettnos_chapters}

\end{document}